\documentclass[10pt,twocolumn,letterpaper]{article}

\usepackage{cvpr}
\usepackage{times}
\usepackage{epsfig}
\usepackage{graphicx}
\usepackage{amsmath}
\usepackage{amssymb}

\usepackage{amsfonts}
\usepackage{algorithm}
\usepackage{algorithmic}
\usepackage{url}
\usepackage{algorithm}
\usepackage{algorithmic}
\usepackage{threeparttable}
\usepackage{amsmath}
\usepackage{longtable}
\usepackage{tabu}
\usepackage{array}
\usepackage{multirow}
\usepackage{rotating} 

\usepackage{amssymb}
\usepackage{latexsym}

\usepackage{lineno,hyperref}
\usepackage{multirow}
\modulolinenumbers[5]
\usepackage{amsmath}
\usepackage{amssymb}
\usepackage{graphicx}
\makeatletter
\newcommand*\bigcdot{\mathpalette\bigcdot@{.5}}
\newcommand*\bigcdot@[2]{\mathbin{\vcenter{\hbox{\scalebox{#2}{$\m@th#1\bullet$}}}}}
\makeatother

\usepackage{url}
\usepackage{xcolor}
\usepackage{authblk}

\usepackage{hyperref}

\cvprfinalcopy % *** Uncomment this line for the final submission

\def\cvprPaperID{****} % *** Enter the CVPR Paper ID here

\begin{document}
\title{RGB-Topography and X-rays Image  Registration for Idiopathic  Scoliosis Children Patient Follow-up}
\author[1,2,*]{Insaf Setitra}
\author[2]{Noureddine Aouaa}
%\fntext[fn1]{isetitra@usthb.dz,isetitra@cerist.dz}
\author[2]{Abdelkrim Meziane}
\author[1]{Afef Benrabia}
\author[3]{Houria Kaced}
\author[3]{Hanene Belabassi}
\author[3]{Sara Ait Ziane}
\author[4]{Nadia Henda Zenati}
\author[4]{Oualid Djekkoune}

\affil[1]{\textit{University of Science and Technology Houari Bouemadiene USTHB, Algiers, Algeria}}
\affil[2]{\textit{Research Center on Scientific and Technical Information, CERIST, Algiers, Algeria}}
\affil[3]{\textit{Physical Medicine and Rehabilitation Service,  Specialized Hospital Bounaama Djilali, CHU Douera, Faculty of Medicine,  Saad Dahleb University, Blida, Algeria}}
\affil[4]{\textit{Center of Development of Advanced Technologies, CDTA, Algiers, Algeria}}
\affil[*]{\small \textit{Corresponding author: isetitra@usthb.dz, isetitra@cerist.dz}}

\maketitle

\begin{abstract}
Children diagnosed with a scoliosis pathology are exposed during their follow up to ionic radiations in each X-rays diagnosis. This exposure can have negative effects on the  patient's health and cause diseases in the adult age. In order to reduce X-rays scanning, recent systems  provide diagnosis of scoliosis patients using solely RGB images. The output of such systems is a set of augmented images and scoliosis related angles. These angles, however, confuse the physicians due to their large number. Moreover, the lack of X-rays scans makes it impossible for the physician to compare RGB  and X-rays images, and decide whether to reduce X-rays exposure or not.
In this work, we exploit both RGB images of scoliosis captured during clinical diagnosis, and X-rays hard copies provided by patients in order to register both images and give a rich comparison of diagnoses. The  work consists in, first, establishing the monomodal (RGB topography of the back) and multimodal (RGB and Xrays) image database, then registering images based on patient landmarks, and finally blending registered images for a visual analysis and follow up by the physician. The proposed registration is based on a rigid transformation that preserves the topology of the patient's back. Parameters of the rigid transformation are estimated using a proposed angle minimization of Cervical vertebra 7,  and Posterior Superior Iliac Spine landmarks of a  source and target diagnoses. Experiments conducted on the constructed database show a better monomodal and multimodal registration  using our proposed method compared to registration using an Equation System Solving based registration.

\end{abstract}
%\linenumbers
\section*{List of abbreviations}
\begin{table}[h!]
\begin{tabular}{ll}
C7   & Cervical vertebra 7            \\
PSIS & Posterior Superior Iliac Spine \\
IC   & Intergluteal Cleft         \\
SC   & Spine Curve                \\
SFSL & Sagittal Frontal Spinal Line   \\
FD   & Frontal Deviations            \\
ESS  & Equation System Solving \\
LSAE & Least Square Affine Estimator \\
RGB & Red, Green, Blue   \\
HSV & Hue, Saturation, Value
\end{tabular}
\end{table}
%\include{introduction}

%\begin{keyword}
%X-rays, image Registration, Monomodal registration, Multimodal registration, Idiopathic  Scoliosis.
%\end{keyword}

%\end{frontmatter}

%\linenumbers

%% main text
\section{Introduction}
Scoliosis (from Greek \textit{skolios} meaning crookedness) is a spinal deformity  where the vertebrae turn relative to each other causing deformations of the spine in the three planes: frontal, axial and sagittal \cite{Janicki2007}. According to Algerian press \cite{Benmansour2014}, scoliosis affects $2$ to $5\%$ of the population. In the  Physical Medicine and Rehabilitation Service of the Specialized Hospital Bounaama Djilali,  scoliosis children patients are diagnosed and followed up using both X-rays and RGB images. The primary reason for which RGB images are used is the will to reduce X-rays exposure for the children, and to automatically extract  scoliosis measurements such as Cobb, Cyphose and Lordose angles  \cite{Safari2019}. \\
Indeed, RGB images are acquired without any risk on the patient's health in contrary to X-rays ionizing radiations. In \cite{Ronckers2010}, a study of cancer mortality in a cohort of 5,573 women with scoliosis and other spine disorders  diagnosed between 1912 and 1965 and  exposed to frequent diagnostic X-rays procedures is performed. The study showed that  after a median follow-up period of 47 years, 1527 women died, including 355 from cancer. Mortality from breast cancer was significantly elevated as compared to other cancers such as lung, cervical and liver cancers. RGB-based  systems, as an alternative to X-rays scanning, allow to follow up scoliosis patients based only on RGB images. The study in \cite{Korvin2014} concludes that, if reinforced by other studies, the analysis with back surface topography parameters may reduce the number of X-rays examinations required to detect increases in the Cobb angle. The study in \cite{Pino2016} expresses also a correlation between topography parameters and Cobb angle. \\

\subsection{Motivation}
An RGB-based system \footnote{$BIOMOD^{TM}$ information \url{https://www.usine-digitale.fr/article/axs-medical-reconstitue-le-dos-en-3d.N281197}.} \footnote{$BIOMOD^{TM}$ official manufacturer \url{https://www.dms.com/fr/biomod-3s}.} is used in the Physical Medicine and Rehabilitation Service of Bounaama Djilali Hospital (CHU Douera) as a reinforcement diagnosis. Children patients are diagnosed using a clinical diagnosis, an analysis of X-rays images brought by the patient, and an RGB diagnosis. Using the RGB-based system by physicians however, does not reduce the number of X-rays scans, nor allows an automatic diagnosis of clinical procedures. Some of the reasons are : (i) the used RGB-based system computes more than $50$ angles, most of them are not understood by the specilaist of Physical therapy and Rehabilitation Medecine (PRM), (ii) the latter could not verify the study in \cite{Korvin2014} since angles computed by the system are not computed in the same manner as they are done in the X-rays scans. (iii) physicians can not register RGB images with X-rays hard copies so as to compare them visually, and (iv) the RGB-based system used does not allow a follow up of patients using several diagnoses.\\
Our solution to the aforementioned  issues consists in exploiting images provided by the RGB-based system and X-rays scans. This exploration allows to construct an RGB and X-rays database. Using the constructed database, we propose a monomodal registration solution. RGB images of several diagnoses  are registered and presented to the physicians which allows a better follow up of patients. A second multimodal solution is proposed in order to register X-rays images with RGB  images. This can allow to better evaluate how RGB  diagnosis can replace or reduce X-rays radiations. 

\subsection{State-of-the-art}
Image registration can be defined as the process that takes as input a target image, and one or several source images, finds corresponding points between target and source image(s), estimates the optimal geometric transformation that best aligns the images and aligns source image(s) using the obtained geometric transformation. The authors of \cite{Maintz1998} classify medical image registration methods according to nine criteria as follows:
\begin{itemize}
\item \textbf{Dimension}: registration performed between 2D-2D, 2D-3D, 3D-3D and time series.
\item \textbf{Nature of registration basis}: extrinsic  (components of the registration are external to the object, for example landmarks fixed on the subject), intrinsic (components of the registration are from the object) and non-image based (positions of the acquisition devices are known).
\item \textbf{Type of transformation}: rigid, affine, projective and curved.
\item \textbf{Domain of transformation}: local or global. If the transformation is performed only on a part of the image, it is local, otherwise it is global. 
\item \textbf{Interaction}: interaction consists of the initialization of the registration process by the choice of areas of interest. It can be interactive, semi-automatic or  automatic.
\item \textbf{Optimization procedure}: either parameters known or searched for by finding an optimum of some function defined on the parameter space.
\item \textbf{Modality}: monomodal (images to be registered belong to the same modality), or multimodal (images to  be  registered  stem  from    different  modalities).
\item \textbf{Subject}: intrasubject (images involved in a registration task are acquired  from  a  single  patient), intersubject (images of different patients or a patient and a model) and Atlas (one  image is  acquired  from  a  single  patient,  and  the  other  image  is constructed  from  an  image  information  database obtained using imaging of many subjects).
\item \textbf{Object}:  the structure concerned by the registration (head, thorax, abdomen, pelvis and perineum, limbs, and spine and vertebrae).
\end{itemize}
Registration is performed following four major steps: (i) features detection, (ii) features alignment, (iii)  registration model estimation and (iv) transformation of the source image. Features can be regions, lines  or points (Harris features, Scale Invariant Feature Transform SIFT, Speedup Robust Features SURF) \cite{Zitova2003}, \cite{Mutneja2015}. Alignment of features can be iconic (computing Intensity Correlation Coefficient), or characteristic (by minimizing the Manhattan distance, the Euclidean kistance or using a Kd tree) \cite{Zitova2003}.\\
In the transformation model estimation, several cases should be thought of, mainly elementary transformations (translation, rotation, rescale, homothety), composed transformations (linear, affine, rigide, projective), and deformable transformations. Given a source image $I_S$ and a target image $I_t$ with their respective matrices $P$ and $P^\prime$  in the homogenous system, the general transformation model is given by :
\begin{equation}
\label{eq.registration1}
\begin{matrix}
P^\prime= M \times P,  &
M=\begin{pmatrix}
a & b & c \\
d & e & f\\
g & h & 1
\end{pmatrix},
\end{matrix}
\end{equation}

\begin{equation}
\label{eq.registration2} 
\begin{matrix}
\begin{pmatrix}
x^\prime \\
y^\prime \\
1
\end{pmatrix}
= \begin{pmatrix}
a & b & c \\
d & e & f\\
g & h & 1
\end{pmatrix} &
\begin{pmatrix}
x \\
y \\
1
\end{pmatrix}
\end{matrix}
\end{equation}

In order to solve Equation \ref{eq.registration1} and get parameters of the registration model using an Equation System Solving ESS, four points for each image are needed. However, in the feature detection step, more features than needed are extracted. Registration becomes an optimization problem which minimizes a cost function. Widespread methods include Gradient Descent, Dynamic Programming, Metaheuristics and Ransac \cite{Nag2017}, \cite{Brown1992}.\\

In \cite{Wang2015} a rough affine transform using Sum of Squared Differences (SSD) similarity measure is applied to register landmarks of two 3D breast MRI. Edge-based non rigid deformation is then estimated using a Normalized Gradient Fields (NGF) similarity measure. Non rigid deformation is suitable to get a deformation field, however, if the goal is to register images without changing their topology, the method cannot be applied. 

Registration of 3D MRI and X-rays mamograms is done in \cite{Garca2019} based on two alternatives; gradient or intensity. In the intensity-based approach, the MRI images are registered using  a Normalized Cross-Correlation (NCC). In the gradient-based approach, the intensity gradients are extracted in the MRI volume and projected into the mammographic space. NCC of the scalar gradients values and gradient correlation of the vectorial gradients are then used to perform the optimization. The study showed that gradient based registration was more suitable for breast MRI registration than intensity based registration.

In \cite{Iglesias2018}, 2D histological and MRI brain images are registered using a deformable probabilistic model.  The model simultaneously solves for registration and synthesis on the target images. Motivation  behind the simultaneity  is that improved registration provides less noisy data for the synthesis, while more accurate synthesis leads to better registration. The framework enables registration and synthesis to iteratively exploit the improvements in the estimates of the other, while considering the uncertainty in each other's parameters. The method is compared with registration using Mutual Information \cite{Polfliet2018} and shows a better result.\\ 
Deep learning approaches for registration gained a particular attention and  have changed the landscape of medical image processing research \cite{Haskins2019}. In \cite{Grewal2020}, authors propose an end-to-end deep learning approach to automatically detect landmarks form two 2D images and match them. In \cite{Yi2020} authors focus their network, consisting of an encoder-decoder, on extracting landmarks from X-rays images of scoliosis patients. Our work does not rely on a deep learning approach since, first, the number of images is limited, and does not allow a proper training, second, landmarks have unified colors, which does not require a deep network, and finally, our approach is dedicated to physician, hence, the method proposed has limited complexity  which allows it to be be used on-line. Though we aim in a future work to use a deep learning approach in order to automatically detect  the spine in the X-rays images of scoliosis.

\subsection{Proposed work}
This work is dedicated to the task of registering topography (RGB) and X-rays images of  scoliosis children patients. For this sake, we follow several steps. First, landmarks positioned on the back of the patient are automatically detected  using thresholding. Landmarks are then registered using an angle minimazation. The registration work is described using the conventional classification of registration \cite{Maintz1998} as follows:
\begin{itemize}
\item \textbf{Dimension}: all images are two-dimensional.
\item \textbf{Nature of registration basis}: extrinsic for RGB images, intrinsic for X-rays images.
\item \textbf{Type of transformation}: rigid, since any deformation generates a loss of scoliosis evolution.
\item \textbf{Domain of transformation}: global.
\item \textbf{Level of interaction}:  automatic.
\item \textbf{Modality}: monomodal for RGB images, multimodal for  X-rays and RGB images.
\item \textbf{Subject}: intrapatient.
\item \textbf{Object}: spine and vertebrae.
\end{itemize}

Our main contributions can be summarized as follows. First, we explore the RGB-based system used in the Physical Medicine and Rehabilitation Service of Bounaama Djilali Hospital, along with its output images in order to construct an initial image database. By analyzing provided images, we define a set of thresholds to segment different areas of the images. These findings make our segmentation methods reproducible for any  RGB-based system users that wish to include a mutli-diagnosis follow up. Our second contribution consists in exploring several methods for scanning X-rays films. The method can be applied to any X-rays images that were not archived in their original numeric format. Our third contribution consists in proposing a simple yet accurate method for registering RGB-RGB and X-rays-RGB images. Accuracy of the method is related to two constraints: (i) RGB images and X-rays images are labeled using unified color landmarks, and (ii)  registration applies  non rigid transformations so as to keep topology of the back. Our method is compared to an Equation System Solving (ESS) and shows a more acceptable result since the use of an ESS does not preserve the topology of the back. The last but not least contribution consists in offering a complete real time framework for physicians who would like to follow up their scoliosis patients both using and RGB-based system  and X-rays images.\\

\section{Image database}

\subsection{Landmarks description}
In our study, we choose images where anatomical landmarks are highlighted and then used for registration. In this section we describe briefly the anatomy of spine and how anatomical landmarks are chosen.\\
The spine, also called  backbone, is a flexible bone column formed by the superposition of $33$ vertebrae. When the spine is healthy, the superposition of the vertebrae in the frontal plane is rectilinear, while  the sagittal plane shows flexibility and elasticity of the spine  \cite{Moses2012}. The spine has two types of physiological curvatures of opposite convexity :
\begin{itemize}
\item Kyphosis: a posterior deviation whose concavity is directed forward.
\item Lordosis: a prior deviation whose concavity is directed backwards.
\end{itemize}
These curvatures divide the spine into five spinal regions \cite{Gesbert2014}:
\begin{itemize}
\item Cervical region, which covers the blow and consists in seven cervical vertebrae C1 to C7 with a lordotic curvature.
\item Thoracic region, which covers the chest and consists in twelve dorsal vertebrae T1 to T12 with a cyphotic curvature.
\item Lumbar region, which covers the lower back and consists in five lumbar vertebrae L1 to L5 with a lordotic curvature.
\item Sacral region, which covers the pelvis and consists in five welded vertebrae from S1 to S5 with a cyphotic curvature.
\item Coccyx region includes four welded vertebrae.
\end{itemize}
\subsection{RGB Topography images}
%\begin{enumerate}
%\item
\subsubsection{Patient preparation}
\begin{figure}[h]
\caption{Landmarks traced by the doctor.}
\label{fig.landmarks}
\centering
\includegraphics[width=0.4\textwidth]{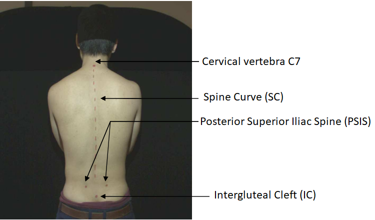}
\end{figure}

The patient, hair tied, naked back, and shoulders not visible, stands in front of a black background (Figure \ref{fig.landmarks}). The doctor then traces, using a random color marker and based on his touching, anatomical landmarks on the patient's back. Anatomical landmarks are:
\begin{itemize}
\item The four principal landmarks:  the $7^{th}$ Cervical vertebra (C7), the two Posterior Superior Iliac Spine (PSIS) and the Intergluteal Cleft (IC) traced with circles. 
\item The Spine Curve (SC) traced with a dotted line. 
\end{itemize}

\subsubsection{Image acquisition}
\begin{figure}[h]
\caption{Image acquisition using the RGB-based System, (a) patient placement (b) projected Moir\'{e} fringes \cite{Bolzinger2017}.}
\label{fig.imageAcquisition}
\centering
(a)\includegraphics[width=0.2\textwidth]{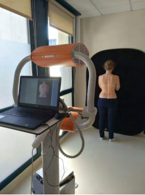}
(b)\includegraphics[width=0.2\textwidth]{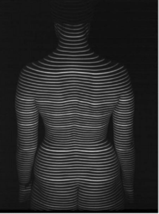}
\end{figure}

The acquisition device is first placed 180 cm facing the back of the patient. Thereafter, the capturing system is calibrated so that a white rectangle projected by the system includes the entire back of the patient \cite{AXS2008}. Images are then acquired using Moir\'{e} fringes technique \cite{Porto2010}. The latter is used to generate a 3D image of the back. Image capturing is illustrated in Figure \ref{fig.imageAcquisition}.

\subsubsection{RGB-based system tracing} 
\begin{figure}[h]
\caption{Automatic detection by the RGB-based system of \textbf{(a)} back surface, \textbf{(b)} landmarks, \textbf{(c)} spine \cite{Cobetto2013}.}
\label{fig.detectedLandmarks}
\centering
\includegraphics[width=0.4\textwidth]{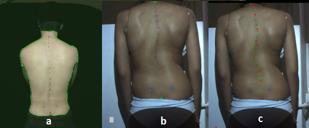}
\end{figure}

After image acquisition, landmarks created by the doctor are detected by the system and traced using a single color (different colors for each landmark but same colors for all patients) as shown in Figure \ref{fig.detectedLandmarks}. The doctor can then adjust this automatic detection. Once adjusted,  the images are generated and stored.

\subsubsection{Augmented images}
\begin{figure}[h]
\caption{Content of an examination (pdf file)  \cite{AXS2008}. The images are provided by the Physical Medicine and Rehabilitation Service of Bounaama Djilali Hospital (CHU Douera). The use of the images is authorized by the  Director of the Service.}
\label{fig.pdfImage}
\centering
\includegraphics[width=0.4\textwidth]{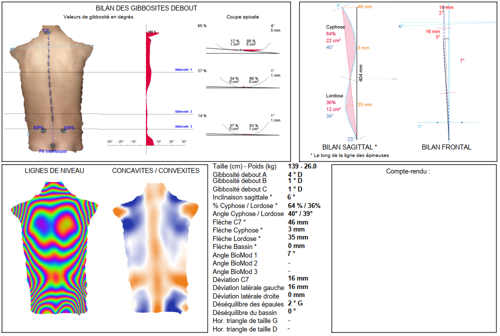}
\end{figure}

\begin{figure}[h]
\caption{Content of a an examination (the viewer zip file). The images are provided by the  Physical Medicine and Rehabilitation Service of Bounaama Djilali Hospital (CHU Douera). The use of the images is authorized by the  Director of the Service.}
\label{fig.viewerContent}
\centering
\includegraphics[width=0.5\textwidth]{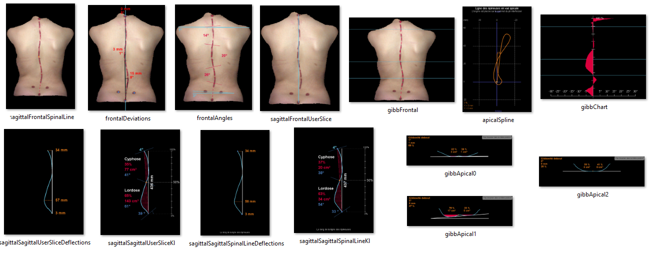}
\end{figure}

For diagnosis, the system computes a set of parameters describing the health of the patient. Parameters are displayed on the acquired images generating augmented images. The doctor can choose what parameters to display and on which plan (frontal, axial or sagittal). The record can then be exported in a pdf (Figure \ref{fig.pdfImage}) or a zip format. The augmented images are stored in the zip file called the viewer (Figure \ref{fig.viewerContent}). The viewer contains 14 augmented images describing the back of the patient such as the hunchback postures, deviations and angles.

\subsubsection{Patient follow-up}
\begin{figure}[h]
\caption{Comparison of two medical examinations provided by the RGB-based system \cite{AXS2008}.}
\label{fig.followup}
\centering
\includegraphics[width=0.4\textwidth]{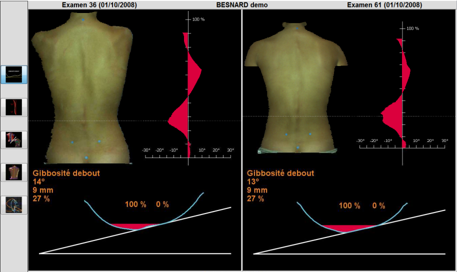}
\end{figure}

Additionally to the generated files, the doctor can compare two exams. The result is a pdf file containing the comparison (Figure \ref{fig.followup}). While comparing two patient examinations is useful, it is not sufficient for the diagnosis of the patient and its follow-up. Registration of the images generated by the system using our method can hence bring an added value to the diagnosis and follow-up of the pathology.

\subsubsection{RGB image sub-database}
Our first RGB image database contains 2 images from each examination of each patient. We choose the following images : Sagittal Frontal Spinal Line (SFSL)  image and Frontal Deviations (FD) image  from the viewer folder of the examination (first and second image in the first row of Figure \ref{fig.viewerContent}). We choose these two images for two main reasons. The first reason is that the SFSL represents the back of the patient as it was seen and marked by the doctor, this image will be displayed after registration for the follow-up. The second image contains augmented landmarks by the RGB-based system. The advantage of these images is the unified color added by the system to enhance the landmarks. Having a unified color can help in segmentation of the landmarks.
%\end{enumerate}

\subsection{X-rays images}
%\begin{enumerate}
%\item 
\subsubsection{X-rays scanning}
\begin{figure}[h]
\caption{X-rays captured with canon IR 1024 scanner. The images are provided by the  Physical Medicine and Rehabilitation Service of Bounaama Djilali Hospital (CHU Douera). The use of the images is authorized by the  Director of the Service.}
\label{fig.X-raysScanned}
\centering
\includegraphics[width=0.15\textwidth]{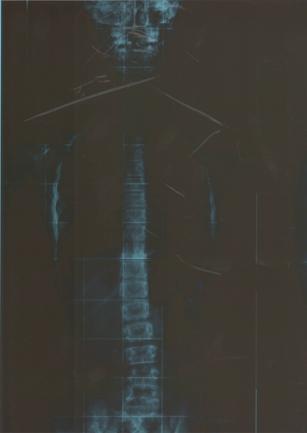}
\end{figure}

\begin{figure}[h]
\caption{X-rays captured with a 7 ppi camera 12 pixels image \textbf{(a)} X-rays film put on a white wall captured with flash, \textbf{(b)} X-rays film put on a window captured with flash, \textbf{(c)} X-rays film put on a window above a white paper captured with flash, \textbf{(d)} X-rays film put on a window above a white paper captured without flash, camera put closer. The images are provided by the  Physical Medicine and Rehabilitation Service of Bounaama Djilali Hospital (CHU Douera). The use of the images is authorized by the  Director of the Service.}
\label{fig.X-raysScanned2}
\centering
(a)\includegraphics[width=0.15\textwidth]{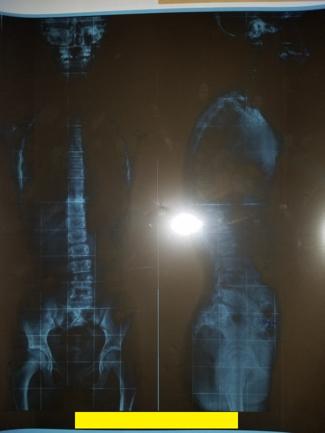}
(b)\includegraphics[width=0.2\textwidth]{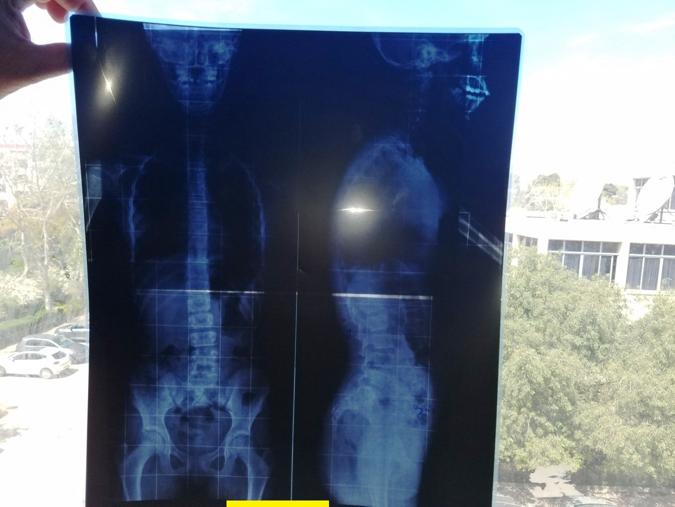}
(c)\includegraphics[width=0.2\textwidth]{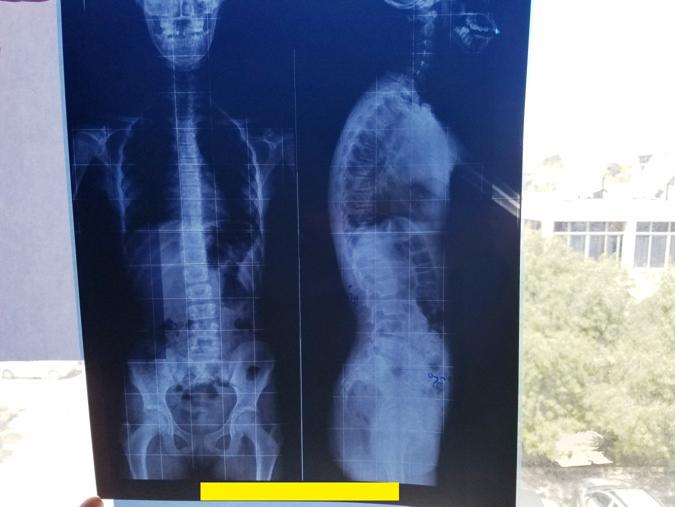}
(d)\includegraphics[width=0.15\textwidth]{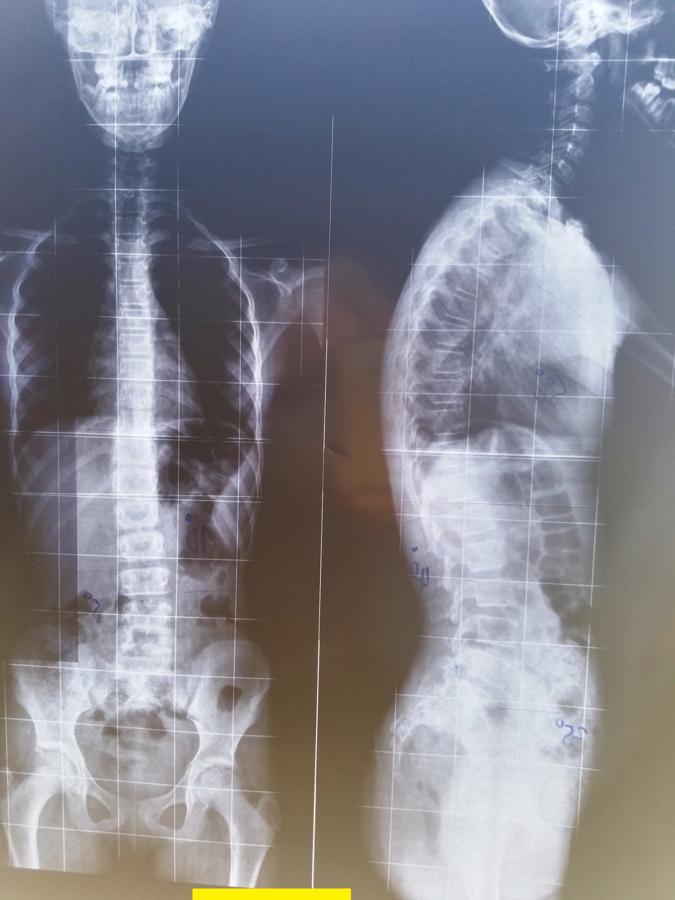}
\end{figure}

\begin{figure}[h]
\caption{X-rays captured with Canon EOS 500D put on a  negatoscope \textbf{(a)} negatoscope used, \textbf{(b)} resulting image after camera calibration.  The images are provided by the  Physical Medicine and Rehabilitation Service of Bounaama Djilali Hospital (CHU Douera). The use of the images is authorized by the  Director of the Service.}
\label{fig.X-raysScanned3}
\centering
(a)\includegraphics[width=0.2\textwidth]{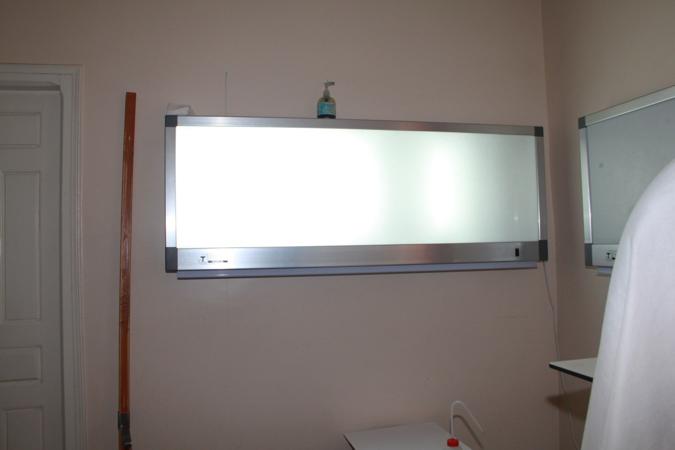}
(b)\includegraphics[width=0.2\textwidth]{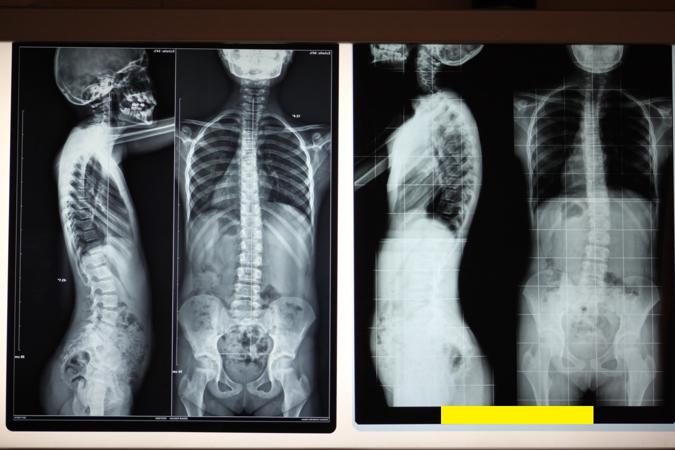}
\end{figure}

Available X-rays images for all patients diagnosed using the RGB-based system are in the form of film photography (hard copies). Hence, in order to register them numerically, several options were exploited. 
The first option, shown in Figure \ref{fig.X-raysScanned} consists in scanning the film using a  canon IR 1024 scanner, the result was unsatisfying. The second option  consists  in using a 7 pixel per inch camera with 12 pixels images ($ 4032 \times 3024$), that captures the films put on a wall, then on a window using daylight with and without white papers put behind the film, and with and without flash. Similarly to the first option, the results were unsatisfying (Figure \ref{fig.X-raysScanned2}). The last option which we chose, was to use a negatoscope and a calibrated Canon EOS 500D camera  to take pictures of the films (Figure \ref{fig.X-raysScanned3}). Resolution of the camera is $ 22,3 \times 14,9 $ mm. Note that for the figures shown in this section, details of the patient are filled with a yellow color. 

%\item 
\subsubsection{X-rays labeling}
\begin{figure}[h]
\caption{Final labeled X-rays image. The images are provided by the  Physical Medicine and Rehabilitation Service of Bounaama Djilali Hospital (CHU Douera). The use of the images is authorized by the  Director of the Service.}
\label{fig.finalX-raysLabeled}
\centering
\includegraphics[width=0.1\textwidth]{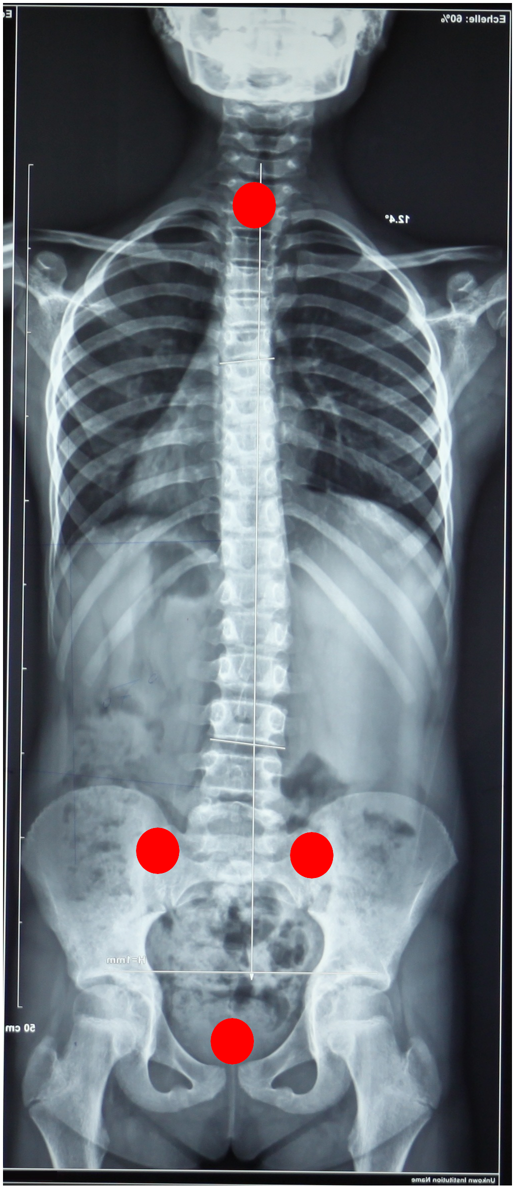}
\end{figure}

In the scanning step, we take pictures of two films put in the negatoscope. The following step consists on cropping the films and removing any zone describing the patient (name, state of birth, address, etc.). For this study, we use only the frontal film, and keep the sagittal one to another study. Labeling of the resulting X-rays image consists on adding landmarks that will help registration with the RGB images. Therefore, we used the same landmarks used by RGB-based system except for the spinal curve. The labeled scans contain the Cervical vertebra C7 (C7) labeled  between Cervical vertebra  C7 and Thoracic vertebra T1, the two Posterior Superior Iliac Spine (PSIS) labeled between Thoracic vertebra T12 and Lumbar vertebra L1, and the Intergluteal Cleft (IC) labeled beteween Lumbar vertebra L5 and Sacral vertebra S1. We use a filled red circle of the same dimension  for all films, even though some films are smaller than others. Figure \ref{fig.finalX-raysLabeled} shows the final X-rays image that will be used for registration.

%\end{enumerate}

\subsection{Final image database}
This study was done on 16 patients. Each patient has 2 to 4 SFSL images, 2 to 4 FD images and 1 to 2 X-rays images.  Table \ref{tab.dabase} shows an overall description of the database. As shown in the table, SFSL and FD images have the same image dimensions for all patients ($ \sigma =0$), X-rays images however differ from one patient to another. Moreover, X-rays images are much larger than SFSL and FD images.

\begin{table*}[!t]
\label{tab.dabase}
\caption{Description of the image database. $\mu$  and $\sigma$ are respectively the mean and the standard deviation of the data, ppi is  the number of pixels per inch.  The use of medical images in this paper is authorized by the Director of the  Physical Medicine and Rehabilitation Service of Bounaama Djilali Hospital (CHU Douera).}
\centering
\begin{tabular}{|c|c|c|c|c|c|}
\hline
\multirow{2}{*}{\textbf{Image}} & \multirow{2}{*}{\textbf{Resolution }} & \multicolumn{2}{c|}{\textbf{Size (KB)}} & \multicolumn{2}{c|}{\textbf{Hight $\times$ Width}} \\ \cline{3-6} 
                                &                                            & $\mu$            & $\sigma$        & $\mu$                     & $\sigma$               \\ \hline
\textbf{SFSLI}                  & 96  ppi                                       & 373,35           & 51,84           & $494\times 755$           & $0\times 0$            \\ \hline
\textbf{FDI}                    & 96   ppi                                      & 373,66           & 51,32           & $494\times 678$           & $0\times 0$            \\ \hline
\textbf{X-rays}                   & 120   ppi                                     & 32674,54         & 66355,00        & $1142,92\times 2493,76$   & $240,60\times 236,12$  \\ \hline
\end{tabular}
\end{table*}

\section{Methods}
\subsection{Image Registration}
Image registration is based on the landmarks Cervical vertebra C7,  the two Posterior Superior Iliac Spine PSIS, the Intergluteal Cleft IC and the Spinal Curve SC described in the previous section. The first step is  to detect automatically these landmarks. The images that need to be seen by pathologists after registration are the  SFSL image since it is the one that does not contain superfluous content that can confuse the pathologist, and the X-rays image. Hence, the aim is to detect automatically landmarks in these images. Since landmarks of the X-rays images were labeled manually using a unified color in the previous step, a simple thresholding allows to obtain these landmarks. Similarly, the detection of the SC in SFSL is straightforward, as the image is augmented by the RBG-based system with a dotted red line (same color for all images). A simple thresholding is therefore applied to get the spinal line. For remaining landmarks the detection in the SFSL is not so obvious. Indeed, colors used by the physician are not similar in all images, and the system does not augment these two landmarks. In order to detect landmarks, we exploit two options: first, use a machine learning approach to classify connected components of the image, and second,  register the images  SFSL and the Frontal Deviations Image (FD) so as to obtain position of the landmarks. Since the landmarks are augmented by the system in the FD image, thresholding can be used to obtain the landmarks. Detection of landmarks is explained in the following.

\subsubsection{Thresholding based C7, SC and IC detection}
The SC is detected by thresolding the SFSL image in the HSV space. Thresholds used reflect the red color added by the RGB-based system to enhance the SC. These global  thresholds were obtained as follows:
\begin{itemize}
\item The SFSL image is first converted from RGB to HSV space.
\item A sliding window is applied in order to get a set of thresholds for each of the three channels.
\item Thresholding using the selected thresholds  is applied for a set of images (of different patients) and the result is analyzed visually  so as to decide on the  thresholds to choose.
\item After fixing the thresholds, a segmentation on all SFSL images of the dataset with global thresholding is applied.
\end{itemize}

After having the SC, the landmarks C7  and IC are simply the highest and lowest points of the detected SC. We define highest as the point that belong to the segmented SC with the smallest index line, and the lowest as the one with the highest index line. The result can be seen in Figure  \ref{fig.PSISregist}.g.

\subsubsection{Connected component based PSIS detection}
\begin{figure}[h]
\caption{Result of PSIS connected components extraction, \textbf{(a)} original SFSL image, \textbf{(b)} Canny contour of (a), \textbf{(c)} image (b) after superfluous content removal, \textbf{(d)} colored connected components to be classified, \textbf{(e)} binarized connected components of PSIS (positive samples, upper image) and non PSIS (negative samples, lower image). The images are provided by the  Physical Medicine and Rehabilitation Service of Bounaama Djilali Hospital (CHU Douera). The use of the images is authorized by the  Director of the Service.}
\label{fig.PSIScc}
\centering
(a)\includegraphics[width=0.1\textwidth]{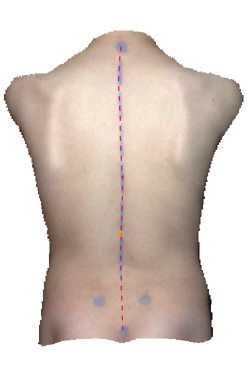}
(b)\includegraphics[width=0.1\textwidth]{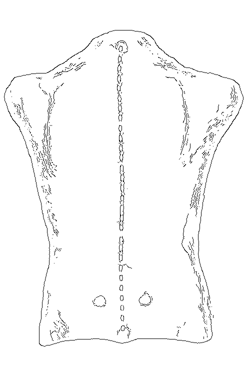}
(c)\includegraphics[width=0.1\textwidth]{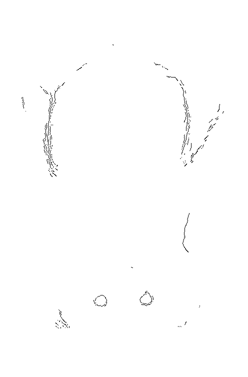}
(d)\includegraphics[width=0.1\textwidth]{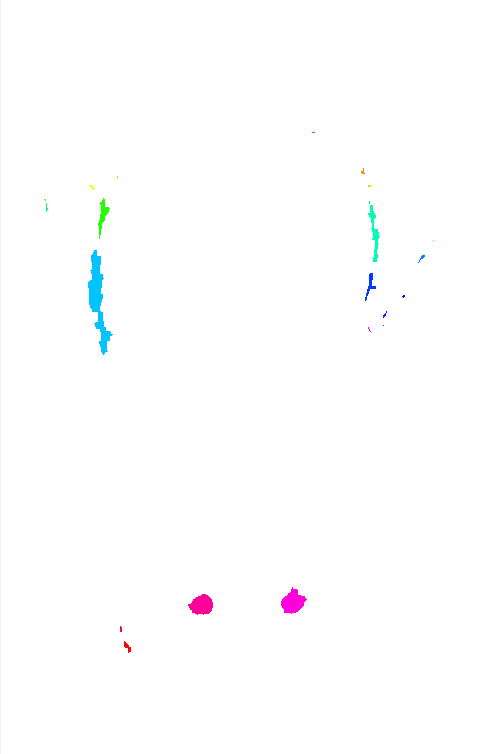}
(e)\includegraphics[width=0.2\textwidth]{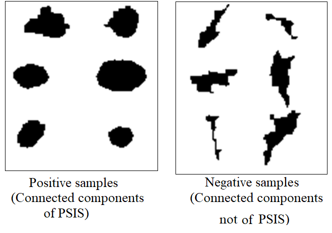}
\end{figure}

\begin{figure}[h]
\caption{Correlation of Aspect Ratio and Circularity features for positive (PSIS connected components) and negative (non PSIS connected components) samples.}
\label{fig.PSIScorrelation}
\centering
\includegraphics[width=0.35\textwidth]{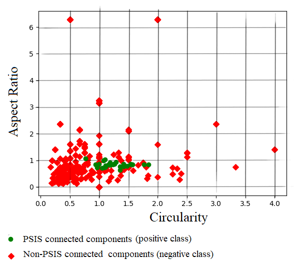}
\end{figure}

In order to detect PSIS landmarks of the SFSL image, we first detect contours using Canny edge detector, then extract connected components, and finally classify the connected components as either a PSIS landmark or not.\\
In order to apply the Canny edge detector which is a hysteresis based contour extractor, we define experimentally the two thresholds. Application of the Canny edge detector, results in both the contour of the PSIS and contour of other uninteresting shapes (the external contour of the shape, the SC, and noise). Removal of the external shape and SC contours is straightdorward. Classifying remaining contours is more complicated.\\
The following step is to apply morphological operations in order to enhance the shape of the PSIS and remove noise. Whilst the PSIS shape is enhanced, still undesired regions remain. \\
The next step is to extract connected components of the cleaned segmented SFSL image. The connected components are manually classified as either PSIS landmark or not PSIS landmark (Figure \ref{fig.PSIScc}.e) in order to construct a training database. Since PSIS landmarks are circular, we extract shape features of the obtained connected components, mainly aspect ratio and circularity. In order to have a preliminary analysis of discrimination power of the chosen shape features, we plot the features with a training data. The result  shown in Figure \ref{fig.PSIScorrelation} shows a low inter-class variance, this finding discouraged us in pursuing a machine learning approach to detect the PSIS landmarks.

\subsubsection{SFSL-FD registration based PSIS detection}
\begin{figure}[h]
\caption{SFSL-FD registration based PSIS detection, \textbf{(a)} original FD image, \textbf{(b)} extracted region of interest of (a), \textbf{(c)} thresholding based PSIS extraction from (a) and (b), \textbf{(d)} extracted contour and PSIS landmarks from (a) colored in green, \textbf{(e)} original SFSL image, \textbf{(f)} extracted region of interest of (e), \textbf{(g)} thresholding based C7, spinal curve and IC landmarks extracted from (f) and (e), \textbf{(h)} extracted contour and PSIS landmarks from (a) after registration with image in (d), \textbf{(i)} blended (a) and (e) before registration, \textbf{(j)} blended (a) and (e) after registration, \textbf{(k)} extracted landmarks from SFSL image after registration. The images are provided by the  Physical Medicine and Rehabilitation Service of Bounaama Djilali Hospital (CHU Douera). The use of the images is authorized by the  Director of the Service.}
\label{fig.PSISregist}
\centering
(a)\includegraphics[width=0.1\textwidth]{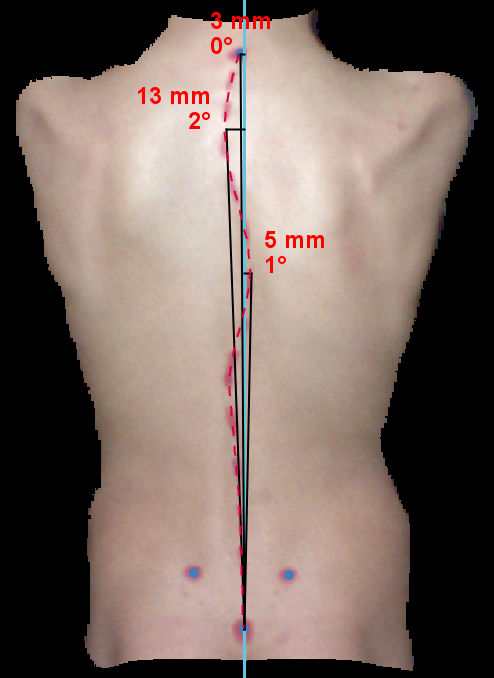}
(b)\includegraphics[width=0.1\textwidth]{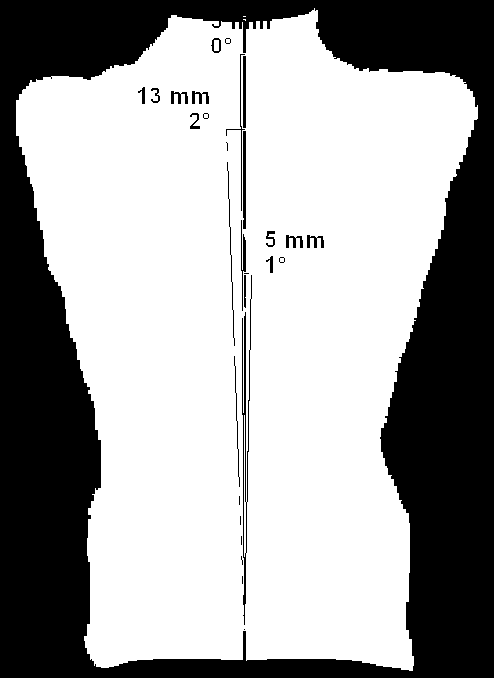}
(c)\includegraphics[width=0.1\textwidth]{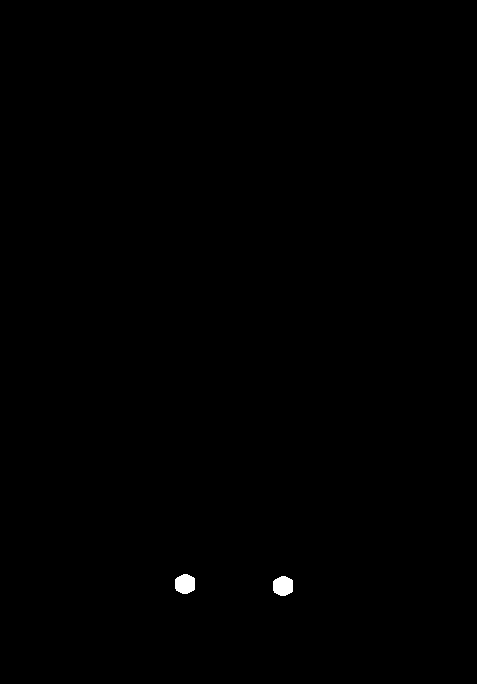}
(d)\includegraphics[width=0.1\textwidth]{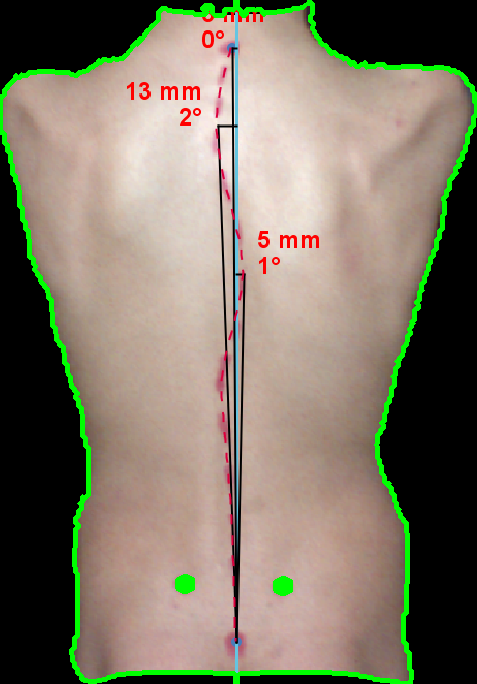}
(e)\includegraphics[width=0.1\textwidth]{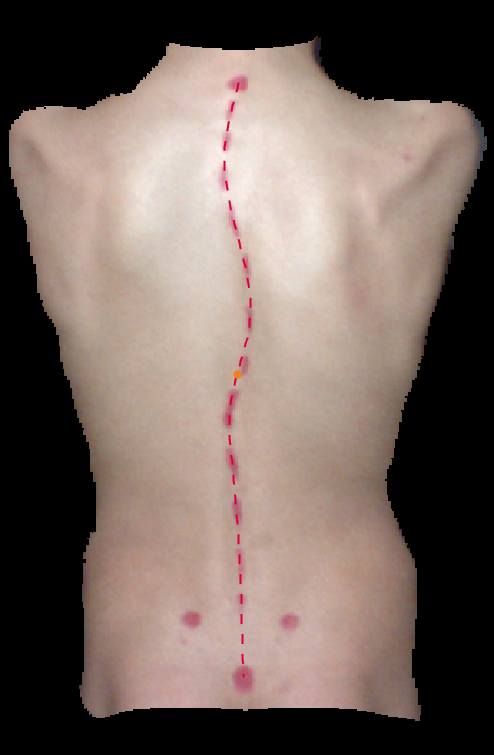}
(f)\includegraphics[width=0.1\textwidth]{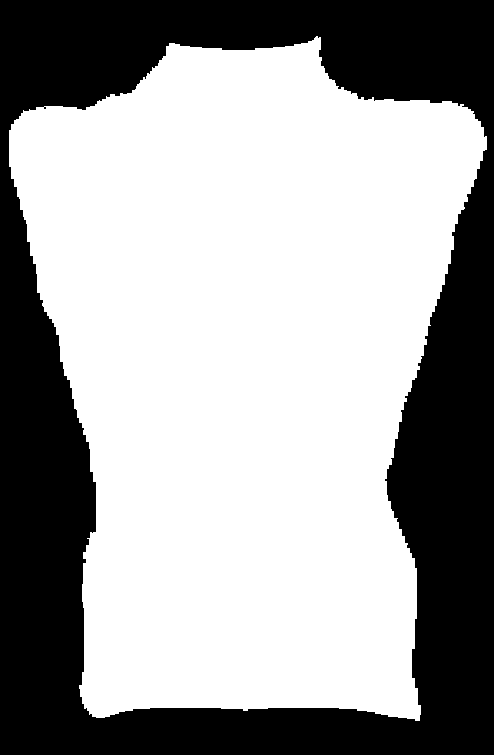}
(g)\includegraphics[width=0.1\textwidth]{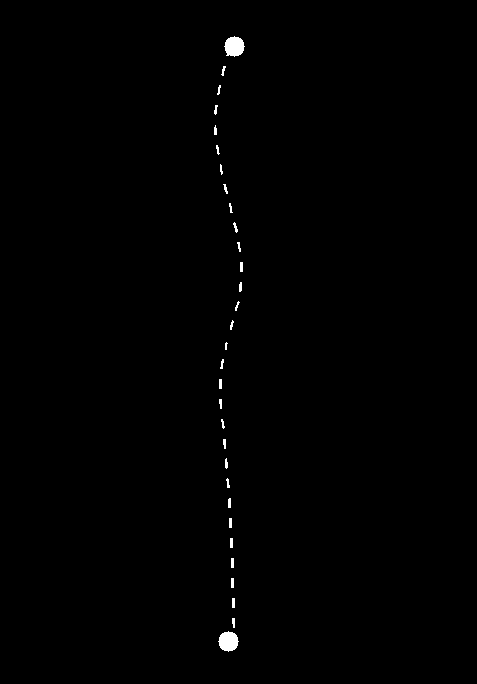}
(h)\includegraphics[width=0.1\textwidth]{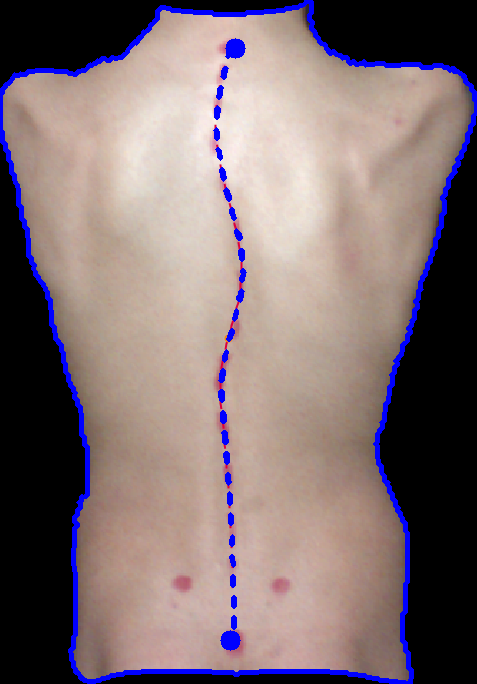}
(i)\includegraphics[width=0.1\textwidth]{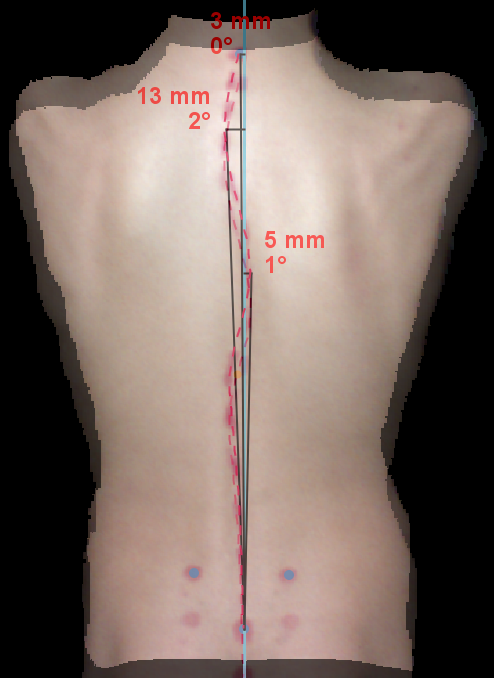}
(j)\includegraphics[width=0.1\textwidth]{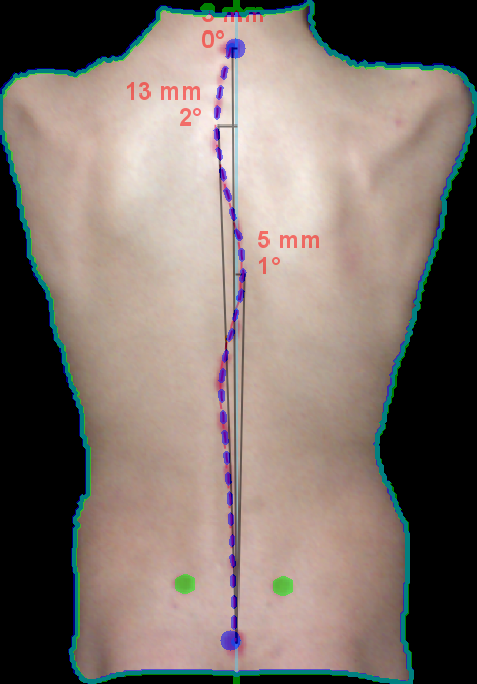}
(k)\includegraphics[width=0.1\textwidth]{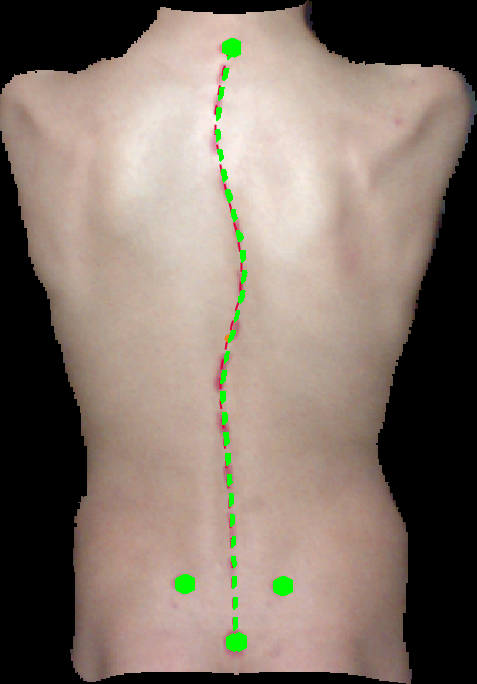}

\end{figure}

As shown in Figure \ref{fig.viewerContent}, the SFSL and FD images are similar, though, the FD image is augmented with labeling (added blue PSIS markers, red spine marker, light blue vertical line, and red numbers and characters). Moreover, according to our investigations, the SFSL image is larger than the FD image. Since only  marked PSIS are needed, we first remove the remaining content, i.e. the blue vertical line and the red scripts. After that, we extract regions of interest from both SFSL and FD images. We consider a region of interest, a region where only the back is visible, i.e. background, blue vertical line and the red scripts removed. After that, the two regions of interests are registered. This is done in order to have the PSIS landmarks detected in the same location for both images. Once the two images aligned, locations of PSIS landmarks are extracted from FD registered image. Each step is explained in the following.\\
For removal of the blue line from the FD images, we apply a global thresholding. Similar steps to the ones followed for C7 and spinal line detection are followed. The image is converted to the HSV space, then the lower and  upper thresholds  are empirically chosen to threshold the FD image and extract the blue line. The blue line is then removed from the original FD image by replacing it with the background (a black background). \\
For script removal, choosing a channel different than the red one results in an image without the red script. Hence, in order to remove the red script, we simply extract the blue channel from the FD image after  blue line removal. From that image, a sub image is subtracted, where another thresholding is applied. Since the image background is black and similar to regions we filled after superfluous content removal, the region of interest is simply pixels different than 0. Extracting the region of interest from the FD image becomes straightforward. Only this last step is applied to the SFSL in order to extract the same region of interest. \\
The following step is to extract the PSIS from the FD region of interest image. Getting the FD region of interest is done by thresholding the image after converting it to a gray level space. Once both regions of interest extracted and coordinates of PSIS detected on the FD image, registration of FD and SFSLT is applied using a nearest neighbor interpolation. After interpolation, the PSIS landmarks are  placed directly on the SFSL image. \\
In Figure \ref{fig.PSISregist} we show the whole SFSL-FD registration based PSIS detection process. Observe in the figure how PSIS landmarks are not aligned before registration (Figure \ref{fig.PSISregist}.j) and are perfectly aligned after registration (Figure \ref{fig.PSISregist}.h) which makes landmarks extraction straightforward. The final image  (Figure \ref{fig.PSISregist}.i) is the one that will be used for registration of multiple patient monomodal and multimodal images.

\subsection{Angle minimization based registration for intra-patient inter-diagnosis images}

After extracting landmarks from SFSL images of examinations of the whole database, the following step consists in registering SFSL images of the same patient for different diagnoses both in a monomolodal and multimodal fashion. To this end this paper proposes a novel angle minimization based registration explained in the following.\\

Let $I_s$ and $I_t$ be the source and target image that we would like to register, the images relates to a patient diagnosis at different times. In the previous step, we extracted from each image the four landmarks : C7, PSIS, and IC. Let the vectors $C_s$, $P^L_s$,$P^R_s$ and $IC_s$ be the points representing the four landmarks of the source image where:
$C_s$ is the vector representing coordinates of the C7 landmark for the source image, $P^L_s$,$P^R_s$  the vectors representing left and right PSIS landmarks respectively for the source image  and $IC_s$  the vector representing $IC$ landmark for the source image. Changing the subscript $s$ to $t$ results in vectors of the target image. Note that for this study, we do not make use of the spine curve SC and leave it for a future work. Note that, by abuse of notation, we use vector and point as the same entities describing the landmarks. \\
In order to register source and target images based on registering the landmarks we use only rigid transformations ( rescale, translation and rotation) since we do not want to change the topology of the back as the scoliosis might be altered.\\

%\begin{itemize}
\textbf{Rescale}: In the first step, we want the source image to be rescaled to the size of the target image. First, we get the rescale factor, then apply a rescale transformatiom. Let $IC_s=\begin{pmatrix} x_s \\ y_s  \end{pmatrix}$, $IC_t=\begin{pmatrix} x_t \\ y_t \end{pmatrix}$,  $P^L_s =\begin{pmatrix}
x^L_s \\ y^L_s  \end{pmatrix}  $, $P^R_s =\begin{pmatrix} x^R_s \\ y^R_s  \end{pmatrix}$, $P^L_t =\begin{pmatrix} x^L_t \\ y^L_t  \end{pmatrix}  $ and $P^R_t =\begin{pmatrix} x^R_t \\ y^R_t  \end{pmatrix}$, and let the rescaled source landmarks be $P_s^{L^\prime} =\begin{pmatrix} x^{L^\prime}_s \\ y^{L^\prime}_s  \end{pmatrix}  $, $P^{R^\prime}_s =\begin{pmatrix} x^{R^\prime}_s \\ y^{R^\prime}_s  \end{pmatrix}$.  Rescaling is performed by first computing the distance between C7 and the middle point of PSIS landmarks for both source and target images, then the scale factor is computed as the ratio between the two distances (source and target), finally, the image is rescaled according to the scale factor.
\begin{Large}
\begin{equation}
\label{eq.getmiddlePoint}
\left\lbrace
 \begin{matrix}
x^M_s=\frac{x^L_s + x^R_s}{2}  \\
y^M_s=\frac{y^L_s + y^R_s}{2} 
\end{matrix}
\right., \qquad \qquad
\left\lbrace
 \begin{matrix}
x^M_t=\frac{x^L_t + x^R_t}{2}  \\
y^M_t=\frac{y^L_t + y^R_t}{2} 
\end{matrix}
\right.
\end{equation}
\end{Large}

\begin{equation}
\label{eq.getEuclideanPoint}
\left\lbrace
\begin{matrix}
d_s=\sqrt{(x_s - x^M_s)^2 + (y_s - y^M_s)^2}  \\
d_t=\sqrt{(x_t - x^M_t)^2 + (y_t - y^M_t)^2}  
\end{matrix}
\right.
\end{equation}

\begin{equation}
\label{eq.getICrescaled}
\begin{pmatrix} x^\prime_s \\ y^\prime_s  \end{pmatrix} = \begin{pmatrix} \frac{d_t}{d_s} & 0 \\ 0 & \frac{d_t}{d_s}  \end{pmatrix}  \begin{pmatrix} x_s \\ y_s  \end{pmatrix}
\end{equation}

\begin{equation}
\label{eq.getPSISLrescaled}
\begin{pmatrix} x^{L^\prime}_s \\ y^{L^\prime}_s  \end{pmatrix} = \begin{pmatrix} \frac{d_t}{d_s} & 0 \\ 0 & \frac{d_t}{d_s}  \end{pmatrix}  \begin{pmatrix} x^L_s \\ y^L_s  \end{pmatrix}
\end{equation}

\begin{equation}
\label{eq.getPSISRrescaled}
\begin{pmatrix} x^{R^\prime}_s \\ y^{R^\prime}_s  \end{pmatrix} = \begin{pmatrix} \frac{d_t}{d_s} & 0 \\ 0 & \frac{d_t}{d_s}  \end{pmatrix}  \begin{pmatrix} x^R_s \\ y^R_s  \end{pmatrix}
\end{equation}
While rescaling landmarks is important for further processing, rescaling the whole image is important for visualization. Hence, following the same principle applied for landmarks, we rescale the whole source image.\\

\textbf{Rotation:} After rescaling the source image, we rotate it in order to register all landmarks while minimizing the distance between them.  In order to find the rotation angle for the source image, we follow several steps. First we bring $C7$ landmark  to the origin $O=(0.0)$ of the orthonormal basis and translate remaining points accordingly, then compute the rotation angle for the source image landmarks so that the distance between the PSIS landmarks is minimal. Minimizing the distance between PSIS of the source and target image is equivalent to simultaneously minimizing the angles  $\measuredangle{P^L_t  C_t  P^L_t}$ and $\measuredangle{P^R_s C_s  P^R_s}$, where $\measuredangle{.}$ is the angle formed by three points centered at the middle point. Let $A= \begin{pmatrix} x_A \\ y_A  \end{pmatrix}$, $B= \begin{pmatrix} x_B \\ y_B  \end{pmatrix}$, $C=\begin{pmatrix} x_C \\ y_C  \end{pmatrix}$ and $D=\begin{pmatrix} x_D \\ y_D  \end{pmatrix}$ be the translated $P^L_t$ ,$P^R_t$, $P^L_s$, and $P^R_s$ respectively. Bringing the landmarks to the origin is done using the following equations:
\begin{equation}
\label{eq.translate1}
\begin{matrix}
\begin{pmatrix} x_A \\ y_A  \end{pmatrix} = \begin{pmatrix} x^L_t - x_t \\ y^L_t - y_t  \end{pmatrix} \qquad, \qquad
\begin{pmatrix} x_B\\ y_B  \end{pmatrix} = \begin{pmatrix} x^R_t - x_t \\ y^R_t - y_t  \end{pmatrix}    
\end{matrix}
\end{equation}

\begin{equation}
\label{eq.translate3}
\begin{matrix}
\begin{pmatrix} x_C \\ y_C  \end{pmatrix} = \begin{pmatrix} x^L_s - x_s \\ y^L_s - y_s  \end{pmatrix}   \qquad, \qquad
\begin{pmatrix} x_D \\ y_D  \end{pmatrix} = \begin{pmatrix} x^R_s - x_s \\ y^R_s - y_s  \end{pmatrix}  
\end{matrix}
\end{equation}

After translation, the angles that need to be minimized simultaneously are $\measuredangle{AOB}$ and $\measuredangle{COD}$. Minimization of the angle is performed following several steps. First, the angle $\theta_A$ between vectors $A$ and $u$ is computed as:  

\begingroup
\Large
\begin{equation}
\label{eq.angleDiff}
\begin{matrix}
\theta_A = \frac{A \bigcdot  u}{\lVert A \lVert \lVert u \lVert} =  \frac{1 \cdot x_A  + 0 \cdot y_A  }{\sqrt{(x_A-1)^2+(y_A-0)^2}}
\end{matrix}
\end{equation}
\endgroup

where $\bigcdot$  and $\lVert \cdot \lVert $ are the dot product and the norm of the vectors respectively and $u$ the unit vector $u=\begin{pmatrix} 1,0 \end{pmatrix}$. Then, the angle $\theta_C$ between vectors $C$ and $u$ is computed as:

\begingroup
\Large
\begin{equation}
\label{eq.angleDiff}
\begin{matrix}
\theta_C = \frac{ C \bigcdot u}{\lVert C \lVert \lVert u \lVert} =  \frac{1 \cdot x_C  + 0 \cdot y_C  }{\sqrt{(x_C-1)^2+(y_C-0)^2}}
\end{matrix}
\end{equation}
\endgroup

The difference between the angles $\theta_A$ and $\theta_C$ is computed using:

\begingroup
\Large
\begin{equation}
\label{eq.angleDiffref1}
\theta_d= \theta_A - \theta_C
\end{equation}
\endgroup

The angle $\theta_{AB}$ between vectors $A$ and $B$ (the angle formed by the left PSIS, C7 and right PSIS of the target image) is computed using:

\begingroup
\Large
\begin{equation}
\label{eq.angleDiff}
\begin{matrix}
\theta_{AB} = \frac{ A \bigcdot  B}{\lVert A \lVert \lVert B \lVert} =  \frac{x_A \cdot x_B  + y_A \cdot y_B  }{\sqrt{(x_A-x_B)^2+(y_A-y_B)^2}}
\end{matrix}
\end{equation}
\endgroup

The angle $\theta_{CD}$ between vectors $C$ and $D$ (the angle formed by the left PSIS, C7 and right PSIS of the source image) is computed using:

\begingroup
\Large
\begin{equation}
\label{eq.angleDiff}
\begin{matrix}
\theta_{CD} = \frac{ C \bigcdot D}{\lVert C \lVert \lVert D \lVert} =  \frac{x_C \cdot x_D  + y_C \cdot y_D  }{\sqrt{(x_C- x_D)^2+(y_C-y_D)^2}}
\end{matrix}
\end{equation}
\endgroup

The rotations that minimize the distance between PSIS landmarks is the one that (i) brings the right PSIS  landmark of the source image to the same coordinate as the right PSIS  landmark of the target image (obtained in Equation \ref{eq.angleDiffref1}), and (ii) brings the two  PSIS landmarks of the source image to a position for which the angle between the left PSIS of the source and target images is equal to the angle formed by the right PSIS landmarks for the target image (angle equal to $\frac{\theta_{AB}- \theta_{CD}}{2}$). The two rotations can be performed in a single rotation for which the angle is the difference between $\theta_d$  and $\frac{\theta_{AB}- \theta_{CD}}{2}$.
This angle called $\theta$ is computed as:

\begingroup
\Large
\begin{equation}
\label{eq.angleDiff}
\theta=  \theta_d - \frac{\theta_{AB}- \theta_{CD}}{2}
\end{equation}
\endgroup

Points are then transformed using the obtained angle $\theta$. The rotation of the translated $P^L_s$ is given by:

\begin{equation}
\label{eq.angleDiff}
\begin{pmatrix} x^\prime_C \\ y^\prime_C \end{pmatrix} = \begin{pmatrix} \cos \theta & -\sin \theta \\ \sin \theta & \cos \theta  \end{pmatrix}  \begin{pmatrix} x_C \\ y_C \end{pmatrix}
\end{equation}

\begin{figure}[h]
\caption{A geometric representation of the different angles involved in registration. \textbf{(a)}. two blended target and source images as seen by the physician, \textbf{(b)}. the blended images in the orthonormal basis, \textbf{(c)}. illustration of landmarks extracted from images of (b), \textbf{(d)}. bringing the landmarks to the origin and and computation of the different angles. The images are provided by the  Physical Medicine and Rehabilitation Service of Bounaama Djilali Hospital (CHU Douera). The use of the images is authorized by the  Director of the Service.}
\label{fig.graphicalIllustr}
\centering
(a)\includegraphics[height=3cm]{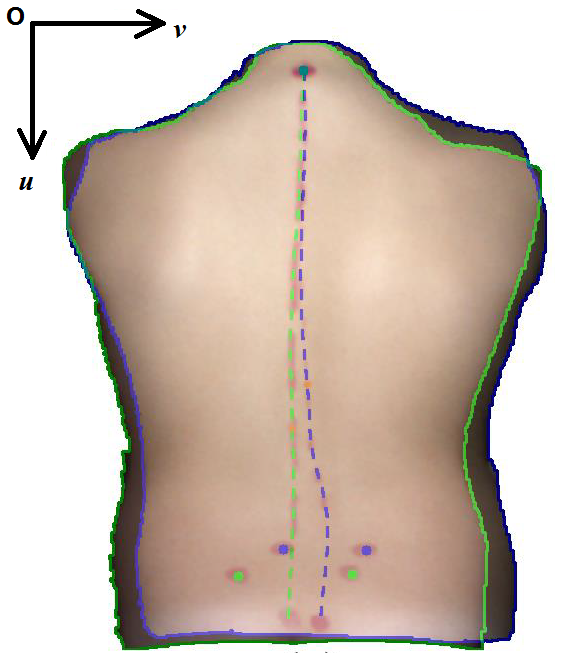}
(b)\includegraphics[width=3cm]{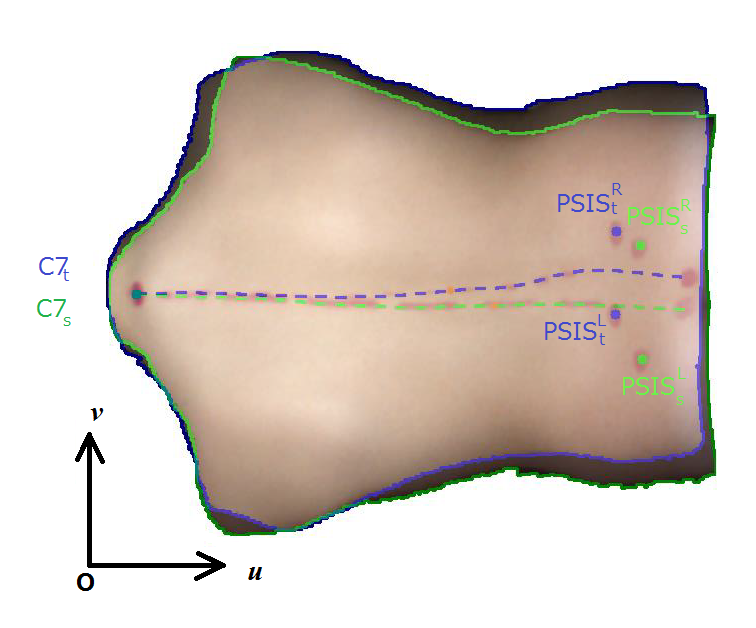}
(c)\includegraphics[width=3cm]{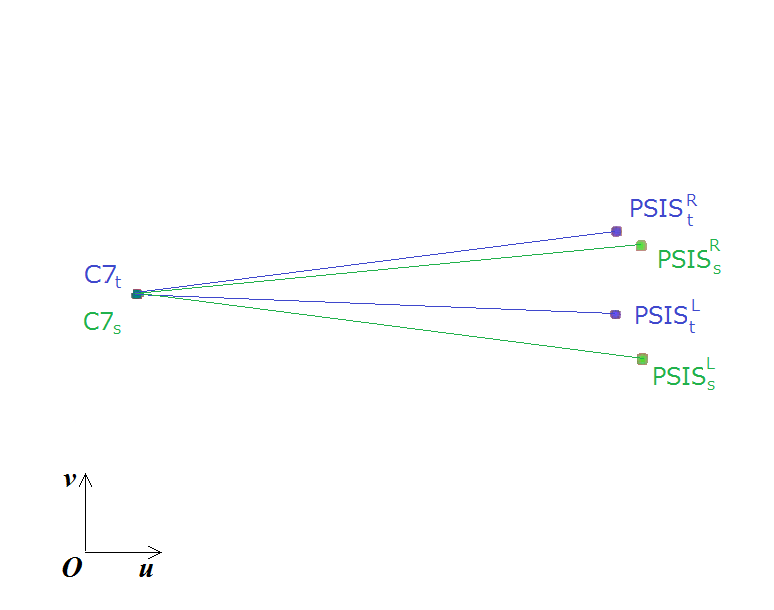}
(d)\includegraphics[width=3cm]{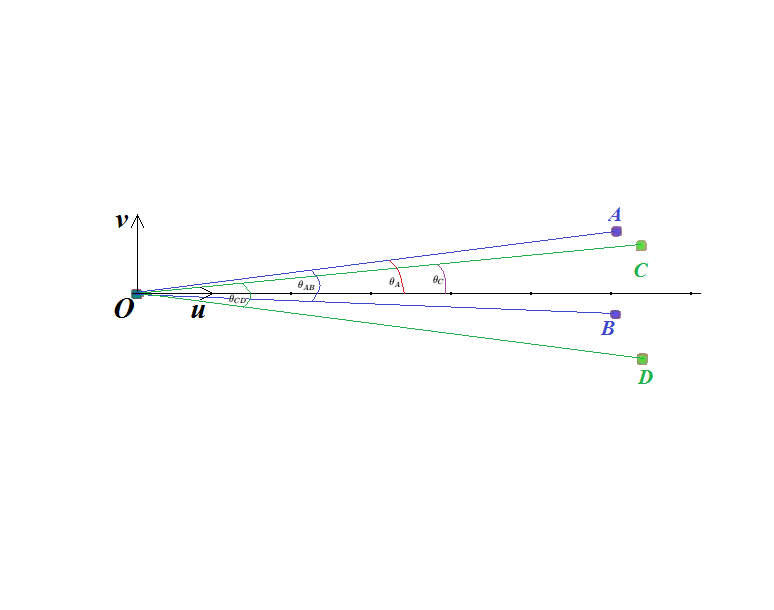}
\end{figure}

According to the obtained angle $\theta$ all points of the image are rotated with respect to the C7 landmark point then brought back to their original coordinates so that to bring back the SFSL image for visualization. A geometric representation of the different angles is illustrated in Figure \ref{fig.graphicalIllustr}.

\textbf{Blending}: Once the images have been  registered (either  monomodal or multimodal), we apply a linear blending in order  to allow visualization of the scoliosis progression. This is done using a blending function :

\begin{equation}
\label{eq.blending}
I = \alpha I_s + (1-\alpha) I_t
\end{equation}

Where $I$ is the resulting image, $0 \leq \alpha \leq 1$ is the user-dependent blending parameter, $I_s$ and $I_t$ are respectively the source and target images that the physician would like to visualize. 

%use https://www.geogebra.org/m/Adc44ZZq for graphics
%\end{itemize}

\section{Experiments}
The method proposed in this paper is adapted to RGB-based system users who wish to have a more advanced follow up for their patients. The method is based on several thresholdings that best suit  for images generated by the system. To the best of our knowledge, no established previous database of RGB-based registration exist so that to compare our result. We compare our method qualitatively to an optimal Least square estimator for affine transformation matrices with uniform scaling, rotation, and translation \cite{palen2016} \footnote{Code for the  optimal Least square estimator for affine transformation matrices with uniform scaling, rotation, and translation \url{https://github.com/axelpale/nudged-py/blob/master/README.rst}.}. %https://github.com/axelpale/nudged-py/blob/master/README.rst
We expose in  this section the obtained results. Results show registration of  RGB SFSL images of different diagnoses for the monomodal registration, and registration of X-rays images with SFSL images for the multimodal registration. We give moreover the set of parameters used for thresholding, along with processing time. The set of parameters allows for a reproduction of the results for Bimod users, and analysis of processing time allows to decide whether to apply the registration in real-time, or offline. Implementations were done using python language and open CV library, and RGB images were obtained using Biomod\footnote{$BIOMOD^{TM}$ information \url{https://www.usine-digitale.fr/article/axs-medical-reconstitue-le-dos-en-3d.N281197}.} \footnote{$BIOMOD^{TM}$ official manufacturer \url{https://www.dms.com/fr/biomod-3s}.} system. 

\subsection{Parameters settings}
In this section we give thresholds for each thresholding used in our method. Table \ref{tab.params} shows parameters used for each thresholding. A pixel is considered as foreground if it is both greater than the lower threshold and lower than the upper threshold in the three channels. Thresholding of RGB images is done in the HSV space while thresholding in X-rays images is done in the RGB space.

\begin{table*}[!t]
\label{tab.params}
\caption{Parameters used for hysteresis image thresholding.}
\centering
\begin{tabular}{|l|l|l|l|l|l|l|}
\hline
\multirow{3}{*}{Segmentation of} & \multicolumn{6}{c|}{RGB image hysteresis thresholding}                   \\ \cline{2-7} 
                                 & \multicolumn{3}{l|}{Lower threshold} & \multicolumn{3}{l|}{Upper threshold} \\ \cline{2-7} 
                                 & H          & S          & V          & H          & S          & V          \\ \hline
Vertical line in FD image        & 97         & 141        & 225        & 97         & 141        & 225        \\ \hline
Spine,C7 and IC in SFSL image    & 12         & 130        & 195        & 180        & 255        & 230        \\ \hline
\multirow{3}{*}{Segmentation of} & \multicolumn{6}{l|}{X-rays image hysteresis thresholding}                     \\ \cline{2-7} 
                                 & \multicolumn{3}{l|}{Lower threshold} & \multicolumn{3}{l|}{Upper thresold}  \\ \cline{2-7} 
                                 & R          & G          & B          & R          & G          & B          \\ \hline
C7, CL and PSIS                  & 255        & 0          & 0          & 255        & 0          & 0          \\ \hline
\end{tabular}
\end{table*}

\subsection{Results}
\begin{figure}[h]
\caption{Result of monomodal registration. \textbf{(a)} target image, \textbf{(b)} source image to be registered, \textbf{(c)} image in (b) registered using Least Square Affine estimator LSAE, \textbf{(d)} image in (b) registered using our method, \textbf{(e)} blending of images in (a) and (b) before registration, \textbf{(f)} blending of images in (a) and (b) after registration using LSAE, \textbf{(g)} blending of images in (a) and (b) after registration using our method. The images are provided by the  Physical Medicine and Rehabilitation Service of Bounaama Djilali Hospital (CHU Douera). The use of the images is authorized by the  Director of the Service.}
\label{fig.result1}
\centering
(a)\includegraphics[width=0.1\textwidth]{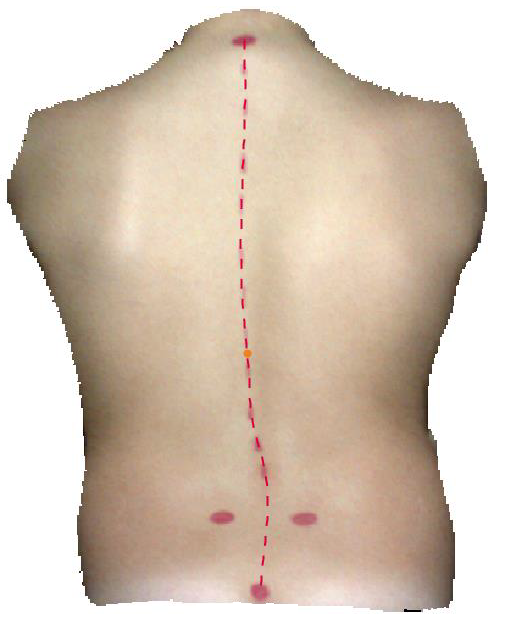}
(b)\includegraphics[width=0.1\textwidth]{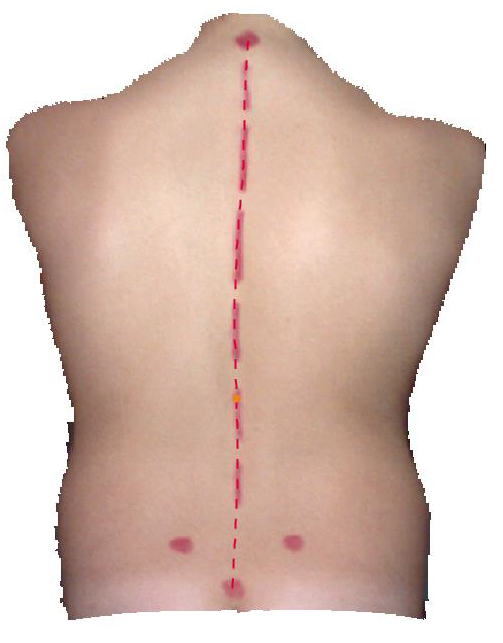}
(c)\includegraphics[width=0.1\textwidth]{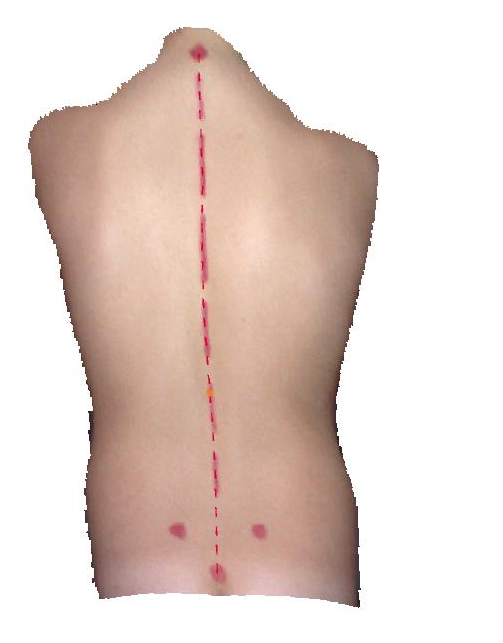}
(d)\includegraphics[width=0.1\textwidth]{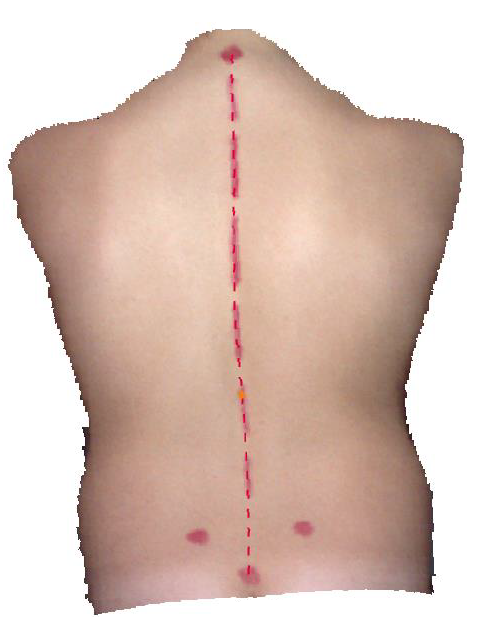}
(e)\includegraphics[width=0.1\textwidth]{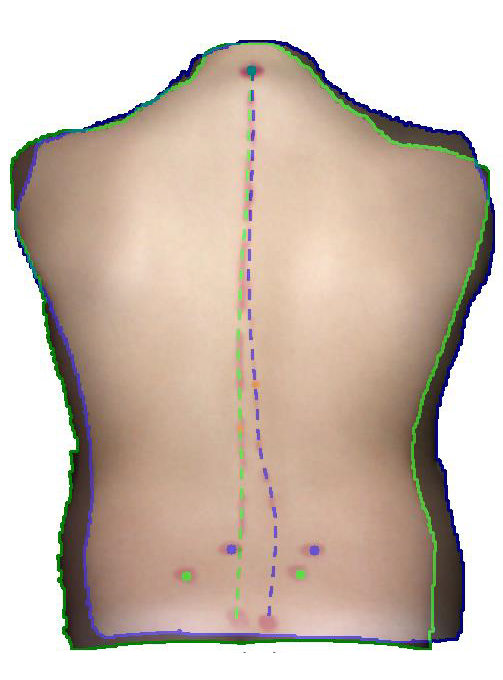}
(f)\includegraphics[width=0.1\textwidth]{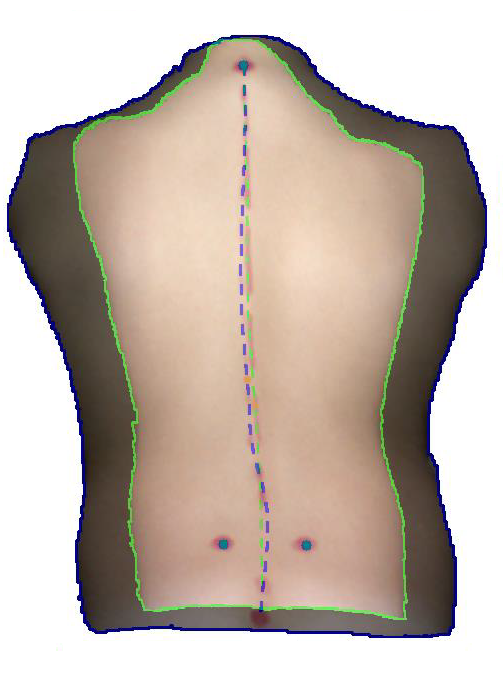}
(g)\includegraphics[width=0.1\textwidth]{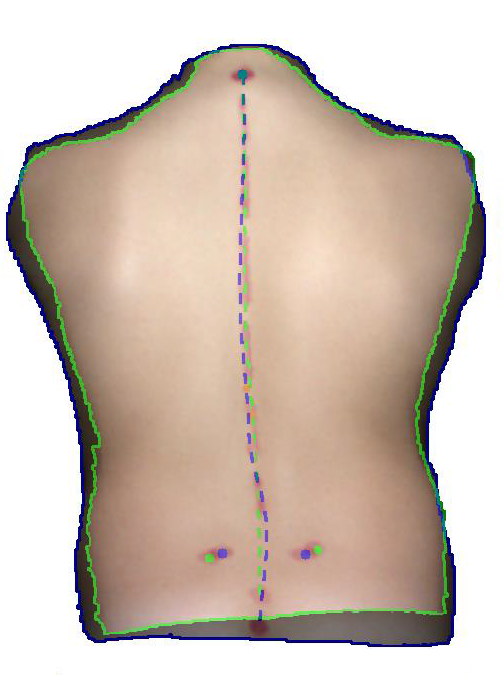}
\end{figure}

For a patient having $N$ diagnosis ($N$ images), number of possible registration is $\frac{N(N-1)}{2}$. In our method, the physician chooses an image as a reference (the target image) and a set of images as sources. Source images will be registered according to the target image. Figure \ref{fig.result1} shows an example of monomodal registration. From the figure we make the following observations:
\begin{itemize}
\item Blending the two diagnosis before registration (Figure \ref{fig.result1}.e) does not allow to correctly compare and/or follow up the patient. The two spinal lines are not aligned which can confuse the physician.
\item Registration based on the Least Square Affine Estimator  LSAE(Figure \ref{fig.result1}.f) shows a perfect alignment of the landmark, however, the back of the patient (the source image) is completely distorted. The spinal line is not reported accurately which might mislead the diagnosis made by the physician.
\item Registration using our method (Figure \ref{fig.result1}.g) shows an acceptable alignment of the landmarks with a minimal distance between the PSIS landmarks for the two diagnoses while preserving the original back topography. Moreover, the registration allows to have an insight of the evolution of the scoliosis. Indeed, the figure shows an improvement of the scoliosis (green spine) compared with the previous diagnosis (blue spine). 
\end{itemize}

\begin{figure}[h]
\caption{Result of multimodal registration. \textbf{(a)} target image, \textbf{(b)} source image to be registered, \textbf{(c)} blending of images in (a) and (b) before registration, \textbf{(d)} blending of images in (a) and (b) after registration using  LSAE, \textbf{(e)} blending of images in (a) and (b) after registration using our method. The images are provided by the  Physical Medicine and Rehabilitation Service of Bounaama Djilali Hospital (CHU Douera). The use of the images is authorized by the  Director of the Service.}
\label{fig.result2}
\centering
(a)\includegraphics[height=0.2\textwidth]{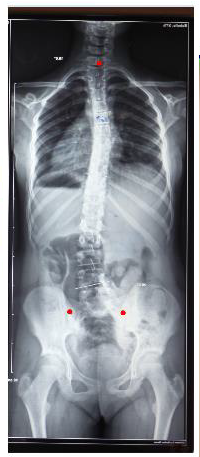}
(b)\includegraphics[height=0.2\textwidth]{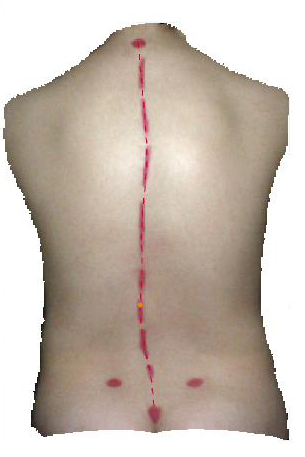}
(c)\includegraphics[height=0.2\textwidth]{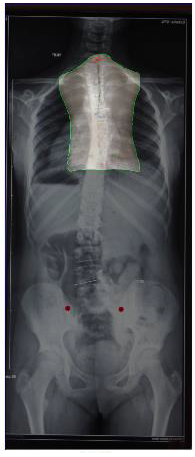}
(d)\includegraphics[height=0.2\textwidth]{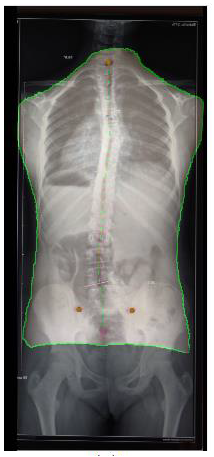}
(e)\includegraphics[height=0.2\textwidth]{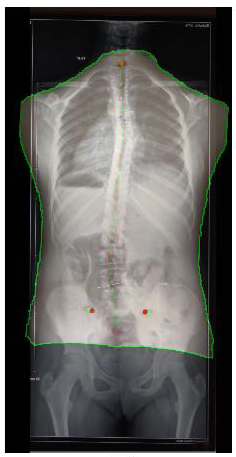}
\end{figure}

Figure \ref{fig.result2} shows an example of multimodal registration. From the figure we make the following observations:
\begin{itemize}
\item Blending of RGB  and X-rays images before registration is not useful for physicians. X-rays images as discussed earlier and shown in Figure  \ref{fig.result2}.c are larger than RGB images. 
\item Registration using LSAE (Figure \ref{fig.result2}.d) does not preserve topology of the back. While landmarks are perfectly aligned, the back looks thinner and the spine curvature is distorted. 
\item Registration using our method (Figure \ref{fig.result2}.e) shows a reasonable result, where the RGB back covers the X-rays skeleton and the distance between  PSIS landmarks is minimized.
\item The spine curvature of the X-rays image is similar to the one in the RGB image. This is mainly due to the correct  labeling made by the physician during the examination. This first observation can encourage the use of RGB-based diagnosis instead of X-rays diagnosis and hence reduce X-rays exposing for the children patients.
\end{itemize}

%\textbf{Performed experiments show robustness of our method in registering different diagnoses with respect of the back's topography which preserves  scoliosis parameters. The main disadvantage of the approach lies in the color based thresholding used to detect PSIS landmarks.This constraint makes the landmarks detection dependent on the Biomod labeling, However, the angle minimization technique can be applied to any . Hence, in a future work, we aim to detect landmarks independently of their colors, which will avoid the manual labeling of the X-rays images.} \textbf{rani nekhallat.... 3awdi had la partie insaf !! it's ok, I can keep things as they are and ignore this part, right?}

\subsection{Computational costs}
Our approach is dedicated to the follow up of scoliosis children patients. Therefore, we need to analyze processing time so that to decide whether the  system based on our method and proposed to the physicians performs registration online or offline. Processing time comprises:   images load, landmarks detection, estimation of the transformation model (parameters of the registration), construction of the transformed images and blending of the different images. These operations go beyond giving only the complexity of our approach, hence, mean  processing time for two diagnoses is presented in Table \ref{tab.processingTime}. As shown in the table, estimation of the transformation model using our method takes the least amount of time and is independent of the size of the images, in contrary to other operations such as loading, registering and blending the images. The longest operation according to the table is the landmarks detection. The operation is even slower for monomodal registration compared to multimodal one. This is because detection of landmarks from an RGB diagnosis and an X-rays diagnosis implies  thresholding and one registration while for detection from two RGB diagnoses implies thresholding and two registrations. Recall that registration is needed in order to detect PSIS landmarks in the SFSL image.

\begin{table*}[!t]
\label{tab.processingTime}
\caption{Mean processing time of our method.}
\centering
\begin{tabular}{l|c|c|}
\cline{2-3}
                                                               & \multicolumn{2}{c|}{\textbf{Mean processing time}} \\ \cline{2-3} 
                                                               & \textbf{Monomodal}           & \textbf{Multimodal}         \\ \hline
\multicolumn{1}{|l|}{\textbf{Images load}}                     & $ 1100 \times e^{-4}$        & $ 1399 \times e^{-3}$       \\ \hline
\multicolumn{1}{|l|}{\textbf{Landmarks detection}}             & $ 3906 \times  e^{-4}$       & $ 2643 \times e^{-3}$       \\ \hline
\multicolumn{1}{|l|}{\textbf{Transformation model estimation}} & $ 24 \times  e^{-4}$         & $ 24 \times e^{-4}$         \\ \hline
\multicolumn{1}{|l|}{\textbf{Transformation of the images}}    & $ 198  \times e^{-4}$        & $ 820 \times e^{-4}$        \\ \hline
\multicolumn{1}{|l|}{\textbf{Blending of the images}}          & $ 673 \times  e^{-4}$        & $ 925 \times e^{-4}$        \\ \hline
\multicolumn{1}{|l|}{\textbf{Total processing time}}           & $ 5889 \times e^{-4}$        & $ 189 \times e^{-2}$        \\ \hline
\end{tabular}
\end{table*}

%\subsection{Scoliosis analysis}
%Here put what physicians think of the registration, Pr Kaced, Dr Belabassi ???
%\subsubsection{Monomodal Registration}
%\subsubsection{Multimodal Registration}

\section{Discussion and Conclusion}
In order to follow up the evolution of scoliosis in children patients, the specialized Hospital Bounaama Djilali uses two image acquisitions, namely, radiography and a  non-invasive RGB-based technology. The latter delivers topography images of the back. Despite the large amount of information given by an RGB-based system, physicians face  difficulties such as the comparison of RGB images of the same patient and the identification of tracing errors of landmarks traced by the physician. The goal of this work is to bring the topographic images of the back to the same geometric reference point so that they can be compared and blended on other topographic or X-rays images. The solution provided is based on landmarks acquisition. Landmarks used in both RGB and X-rays images are the  Cervical vertebra 7, and Posterior Superior Iliac Spine PSIS landmarks. Detection of the landmarks in the RGB images is done using thresholding and registration of the augmented images generated by the RGB-based system. For X-rays images, landmarks were marked manually by physicians. For the registration phase, a transformation model based on rotation, homothety and translation was proposed in order to preserve the topology of the back. For rotation parameters extraction, an angle minimization between PSIS landmarks of two diagnoses is proposed. Finally, blending of the registered images allowed to better observe  differences between monomodal and multimodal registered diagnoses. Processing time analysed in the work showed an adequacy of the method for real time processing. \\

Main limitations of the work comes from  its reliance on color thresholding, which, in the one hand, makes it dependent on RGB images provided by the used RGB-based system, and on the other hand requires manual labeling of the X-rays images. As a solution to this problem, and as a future work, we would like to use a machine learning, or a deep learning technique such as in \cite{Wu2017} to automatically detect the spinal line and the different landmarks in both RGB and X-rays images. As another perspective, we would like to further analyze the obtained results and decide on the frequency of X-rays diagnosis according to intrinsic parameters such as the age of the patient or the Cobb angle. Moreover, since the RGB-based also provides  3D images of the back's topography, we aim in a future work to register  3D RGB images, reconstruct a  3D X-rays model of the back based on the frontal and sagittal X-rays images, and register all images for a better follow up of the physician.

\section*{Acknowledgment}
Authors would like to express their gratitude to  Dr. Djamal Belazzougui for his thorough remarks and guidance that helped in writing the paper. Authors would like also to thank Mr.  Riad Ouyahia and Mr. Madjid Boudina, from Audio-visual Department of Cerist for their valuable help in capturing the X-rays images.

%%Harvard
%\bibliographystyle{model2-names}\biboptions{authoryear}
%\bibliography{bibliography}

{\small
\bibliographystyle{ieee}
\bibliography{bibliography}
}

\end{document}